\documentstyle{ichep}
\include{mytex}
\makeatletter
\newdimen\normalarrayskip              
\newdimen\minarrayskip                 
\normalarrayskip\baselineskip
\minarrayskip\jot
\newif\ifold             \oldfalse
\newif\ifdisplayarray    \displayarraytrue
\newif\ifbigarray        \bigarraytrue
\def\arraymode{\ifold\relax\else\ifdisplayarray\displaystyle\else\relax\fi\fi}
\def\eqnumphantom{\phantom{(\theequation)}}     
\def\@arrayskip{\ifold\baselineskip\z@\lineskip\z@\else\ifbigarray
     \baselineskip\normalarrayskip\lineskip\minarrayskip
     \else
     \baselineskip\z@\lineskip\z@\fi\fi}
\def\@arrayclassz{\ifcase \@lastchclass \@acolampacol \or
\@ampacol \or \or \or \@addamp \or
   \@acolampacol \or \@firstampfalse \@acol \fi
\edef\@preamble{\@preamble
  \ifcase \@chnum
     \hfil$\relax\arraymode\@sharp$\hfil
     \or $\relax\arraymode\@sharp$\hfil
     \or \hfil$\relax\arraymode\@sharp$\fi}}
\def\@array[#1]#2{\setbox\@arstrutbox=\hbox{\vrule
     height\arraystretch \ht\strutbox
     depth\arraystretch \dp\strutbox
     width\z@}\@mkpream{#2}\edef\@preamble{\halign \noexpand\@halignto
\bgroup \tabskip\z@ \@arstrut \@preamble \tabskip\z@ \cr}%
\let\@startpbox\@@startpbox \let\@endpbox\@@endpbox
  \if #1t\vtop \else \if#1b\vbox \else \vcenter \fi\fi
  \bgroup \let\par\relax
  \let\@sharp##\let\protect\relax
  \@arrayskip\@preamble}
\def\eqnarray{\stepcounter{equation}%
              \let\@currentlabel=\theequation
              \global\@eqnswtrue
              \global\@eqcnt\z@
              \tabskip\@centering
              \let\\=\@eqncr
              $$%
 \halign to \displaywidth\bgroup
    \eqnumphantom\@eqnsel\hskip\@centering
    $\displaystyle \tabskip\z@ {##}$%
    &\global\@eqcnt\@ne \hskip 2\arraycolsep
         \hfil$\arraymode{##}$\hfil
    &\global\@eqcnt\tw@ \hskip 2\arraycolsep
         $\displaystyle\tabskip\z@{##}$\hfil
         \tabskip\@centering
    &{##}\tabskip\z@\cr}
\newenvironment{marray}{\begin{equation}\begin{array}}%
{\end{array}\end{equation}}
\newenvironment{carray}{\begin{equation}\begin{array}{rcl}}%
{\end{array}\end{equation}}
\def\be{\@ifnextchar[{\def\ee{\end{equation}}\begin{equation}\l@b}%
{\def\ee{$$}$$}}
\def\l@b[#1]{\label{#1}}
\def\ba{\@ifnextchar[{\def\ee{\end{carray}}\begin{carray}\l@b}%
{\def\ee{\end{array}$$}$$\begin{array}{rcl}}}
\def\barray#1{\@ifnextchar[{\def\ee{\end{marray}}\begin{marray}{#1}\l@b}%
{\def\ee{\end{array}$$}$$\begin{array}{#1}}}
\def\herring{\@ifnextchar[{\@herring}{\@herring[\vcenter]}}
\def\@herring[#1]#2{\begingroup
\def\*{\\ \>}
\topsep0pt
\partopsep0pt
\def\tabbing{\lineskip\jot \lineskiplimit\jot
     \let\>\@rtab\let\<\@ltab\let\=\@settab
     \let\+\@tabplus\let\-\@tabminus\let\`\@tabrj\let\'\@tablab
     \let\\=\@tabcr
     \global\@hightab\@firsttab
     \global\@nxttabmar\@firsttab
     \dimen\@firsttab\@totalleftmargin
     \global\@tabpush0 \global\@rjfieldfalse
     \trivlist \item[]\if@minipage\else\vskip\parskip\fi
     \setbox\@tabfbox\hbox{\rlap{\indent\hskip\@totalleftmargin
       \the\everypar}}\def\@itemfudge{\box\@tabfbox}\@startline\ignorespaces}
\def\@startfield{\global\setbox\@curfield\hbox
                    \bgroup$\displaystyle}%
\def\@stopfield{$\egroup}%
#1{\begin{tabbing}#2\end{tabbing}}\endgroup}

\def\eq#1{(\ref{#1})}
\def\theequation{\thesection.\arabic{equation}}
\@addtoreset{equation}{section}%
\def\@citex[#1]#2{\if@filesw\immediate\write\@auxout{\string\citation{#2}}\fi
  \def\@citea{}\@cite{\@for\@citeb:=#2\do
    {\@citea\def\@citea{,\penalty\@m}\@ifundefined  
       {b@\@citeb}{{\bf ?}\@warning
       {Citation `\@citeb' on page \thepage \space undefined}}%
\hbox{\csname b@\@citeb\endcsname}}}{#1}}
\def\@sect#1#2#3#4#5#6[#7]#8{\ifnum #2>\c@secnumdepth
     \def\@svsec{}\else
     \refstepcounter{#1}\edef\@svsec{\csname the#1\endcsname.%
     \hskip 0.8em }\fi
     \@tempskipa #5\relax
      \ifdim \@tempskipa>\z@
        \begingroup #6\relax
          \@hangfrom{\hskip #3\relax\@svsec}{\interlinepenalty \@M #8\par}%
        \endgroup
       \csname #1mark\endcsname{#7}\addcontentsline
         {toc}{#1}{\ifnum #2>\c@secnumdepth \else
                      \protect\numberline{\csname the#1\endcsname}\fi
                    #7}\else
        \def\@svsechd{#6\hskip #3\@svsec #8\csname #1mark\endcsname
                      {#7}\addcontentsline
                           {toc}{#1}{\ifnum #2>\c@secnumdepth \else
                             \protect\numberline{\csname the#1\endcsname}\fi
                       #7}}\fi
     \@xsect{#5}}

\makeatother

\begin{document}
\newcommand{\dis}{\displaystyle}
\title{{\LARGE{\bf Active Sterile Neutrino Conversions in a Supernova
with Random Magnetic Fields}}}
\author{ {\large {\bf S. Pastor, V. Semikoz and Jos\'e W. F. Valle $^{\dag}$}}}
\address{\dag\
Departament de F\'{\i}sica Te\'{o}rica, Universitat de Val\`{e}ncia
and \\ Instituto de F\'{\i}sica Corpuscular - C.S.I.C.,
E-46100 Burjassot, Val\`encia, SPAIN         }
\abstract
{Large enough random magnetic fields may affect in an
important way neutrino conversion rates, even in the
case where neutrinos have zero transition magnetic
moments. We consider their effect in the case of active
to sterile \neu conversions in a supernova and show that
for KeV neutrino masses these limits may overcome those
derived for the case of zero magnetic field.}

\twocolumn[\maketitle]

\section{Introduction}
\hspace{0.5cm}

There have been several hints for nonzero \neu masses
from astrophysical and comological observations which,
taken altogether, point towards a class of extensions of the standard
model that contain a light sterile neutrino \cite{pa7}.

So far the most stringent constraints for the neutrino
mass matrix including a fourth neutrino species, $\nu_s$,
come from the nucleosynthesis bound on the maximum
number of extra neutrino species that can reach thermal
equilibrium before nucleosynthesis and change
the primordially produced helium abundance \cite{BBNUTAU}.
This has been widely discussed in the case of the
early Universe hot plasma without magnetic field,
as well as recently for the case of a large random
magnetic field (r.m.f.) \cite{SemikozValle}.

Stringent constraints on the active to sterile \neu
oscillation parameters have been derived for the case
of supernovae with zero magnetic field in ref. \cite{Maalampi}.
Here we summarize the results of ref. \cite{pastor}
on the effect that a large supernova r.m.f. has on
the active sterile \neu conversions. This was
motivated by a recent paper \cite{Thomson} which
showed that magnetic fields as strong as $10^{14}$
to $10^{16}$ Gauss might be generated during the first
seconds of neutrino emission inside a supernova core.
If such field is generated after collapse it
could be viewed as the random superposition of many
small dipoles with size $L_0 \sim 1$ Km \cite{Thomson}.
Although the magnetic field in different domains is
randomly aligned relative to the neutrino propagation
direction, the neutrino conversion probabilities
depend on the mean-squared random field via a squared
magnetization value, leading therefore to nonvanishing
averages over the magnetic field distribution.

The effect which we have found is of more
general validity than that which could be ascribed
to nonzero magnetic (transition) moments, as it would exist
even these are negligible, as expected in the simplest
extensions of the standard model.

\section{Active-sterile neutrino conversions in the
presence of a large r.m.f. }
\hspace{0.5cm}

The equation of motion for a system of one active
and one sterile neutrinos propagating in the presence of a large r.m.f. can
be written in terms of weak eigenstates, as
\barray{ll}[v1.20]
i \frac{d}{dt} \Bigl (\matrix{ \nu_a \\ \nu_s }\Bigr ) =
\Bigl (
\matrix{ H_{aa} & H_{as} \\
 H_{as} & H_{ss} }\Bigr ) \Bigl (\matrix{ \nu_a \\
\nu_s }\Bigr ),
\ee
where the quantities in the evolution hamiltonian
are given as
\bea
H_{aa} = (c^2m^2_1 + s^2m^2_2)/2q + V_{as} + A_{as} \\ \nonumber
H_{as} = c s \Delta \\ \nonumber
H_{ss} = (s^2m^2_1 + c^2m^2_2)/2q
\eea
and we have denoted by $V_{as}$ and $A_{as}$ the vector and
axial parts of the neutrino potential that will describe the
active to sterile conversions, given as
\beq
V_{as} \approx 4\times 10^{-6}\rho_{14}(3Y_e + 4Y_{\ne} - 1) \rm{MeV},
\label{f6}
\eeq
\beq
\label{Axial}
A_{as} (q, B) = V_{axial} \frac{q_z}{q}
\eeq
where the term $V_{axial}$ is produced by the {\sl mean axial current} and
is proportional to the magnetization of the plasma
in the external magnetic field, assumed to be pointed
along the z-direction inside a given domain.
In the above equations q is the \neu momentum,
$m_1$ and $m_2$ are the masses of the \neus, $\theta$
is their mixing angle and we use the standard definitions
$\Delta = \Delta m^2/2q$;
$\Delta m^2 = m^2_2 - m^2_1$;
$c = \cos \theta$, and $s = \sin \theta $. The Y's
denote particle abundances and $\rho_{14}$ denotes
the density in units of $10^{14}$ g/cc.

Notice that, although majorana \neus could have nonzero
transition magnetic moments \cite{BFD}, we have neglected
them in our present discussion. As we will see, even in
this case, there may be a large effect of the magnetic
field on the conversion rates.

{}From \eq{v1.20} one can easily obtain the probability
$P_{\nu_a \to \nu_s} (t)$ for converting the active
neutrinos $\nu_a$ emitted by the supernova into the sterile
neutrinos, $\nu_s$. In a strong random magnetic field one can write
\be[v1.30]
P_{\nu_a \ra \nu_s} (B,t) \approx
\frac{\Delta^2 \sin^2 2\theta}{2\Delta^2_m}\Bigl (1 -
\exp (-\Delta^2_mt/2\Gamma )\Bigr ),
\ee
which describes the aperiodic behaviour of the active
to sterile \neu conversion. The relaxation time defined as
\be[relax]
t_{relax} = 2\Gamma/\Delta_m^2 = \VEV{\Delta^2_B} L_0/\Delta_m^2
\ee
depends on the mean squared magnetic field parameter
\beq
\langle \Delta^2_B \rangle^{1/2} = {\abs{\mu_{eff}}
\VEV{{\bf B}^2}^{1/2}\over \sqrt{3}},
\label{f1}
\eeq
where $\mu_{eff}$ is defined in ref. \cite{pastor}
and $L_0$ is the domain size where the magnetic field
is taken as uniform and constant. In \eq{v1.30} the
quantity $\Delta_m$
\beq
\Delta_m = [(V_{as} - \Delta \cos 2\theta )^2 + \Delta^2 \sin^2 2\theta]^{1/2}
\label{f5}
\eeq
is the standard oscillation frequency in the supernova medium
\cite{MSW}.

Note that \eq{v1.30} is valid when $\Gamma \gg \Delta_m$ is fulfilled
and this holds in the case of a very strong r.m.s. magnetic
field \O($10^{14}$ - $10^{16}$) Gauss.

The relaxation time in \eq{relax} can be much larger
than the mean active \neu collision time
$t_{coll} = \Gamma_a(B \neq 0)^{-1}$. In order to see
this we have used the estimate \cite{SchrammB},
\beq
\Gamma_a (B \neq 0) \lsim 2 B_{14} \Gamma_a (B = 0)
\label{B14}
\eeq
where $B_{14}$ denotes the magnetic field strength in units of
$10^{14}$ Gauss.
As we can see this collision rate could be larger than
$\Gamma_a (B = 0)$ by a factor $2 B_{14}$.
This allows us, following ref. \cite{SemikozValle}, to average
\eq{v1.30} over collisions so as to obtain
\beq
\VEV{P_{\nu_a \ra \nu_s}  (B)} = \frac{\Delta^2 \sin^2 2 \theta }
 {\VEV{\Delta^2_B} 4 \Gamma_a L_0} \equiv \frac{\sin^2 2\theta_B}{2}.
\label{f11}
\eeq
where we define the mixing angle in the presence of the
magnetic field via
\beq
\sin^2 2 \theta_B = \frac{\Delta^2 \sin^2 2\theta}
{2 \VEV{\Delta^2_B} \Gamma_a L_0}
=  \frac{x}{2} \sin^2 2\theta_m~,
\label{sinmag}
\eeq
in analogy with the case of zero magnetic field, where
$P_{\nu_a \ra \nu_s} (B \ra 0) = \sin^22\theta_m/2$.
The parameter $x$ is defined as
\beq
\label{argument}
x = \Delta^2_m/ 2 \Gamma \Gamma_a (B \neq 0)
\eeq

\section{Supernova Constraints}
\hspace{0.5cm}

There are two ways to place constraints on \neu oscillation
parameters using astrophysical criteria, depending on the relative
value of the effective sterile neutrino effective mean free path
$l_s \equiv \Gamma_s^{-1} \equiv [P (\nu_a \ra \nu_s) \Gamma_a]^{-1}$
and the core radius $R_{core}$.
If the trapping condition $l_s \leq R_{core}$ is fulfilled,
the \ns are in thermodynamical equilibrium with the medium and,
due to the Stefan-Boltzman law, the ratio of the sterile neutrino
luminosity to that of the ordinary neutrinos,
\beq
\frac{Q_s}{Q_a} \simeq \Bigl (\frac{T(R_s)}{T(R_a)}\Bigr )^4\Bigl (
\frac{R_s}{R_a}\Bigr )^2 \simeq \Bigl (\frac{\Gamma_a}{\Gamma_s}\Bigr )^
{1/2}
= \left(\frac{\sin^2 2\theta_m}{2}\right)^{-1/2} \: ,
\label{limit1}
\eeq
does not depend on $\Gamma_a$.
In this first regime one considers surface thermal neutrino
emission and sets the conservative limit $(Q_s/Q_a)_{max} \gsim 10$
in order to obtain the excluded region of \neu parameters, valid
for $\Delta m^2 \gsim$ KeV$^2$ \cite{Maalampi}
\footnote{The cosmological arguments that forbid
\neu masses in the KeV range or above are not
applicable in models with unstable \neus
that decay via majoron emission  \cite{pa7}. }
\beq
\label{}
\sin^2 2\theta_m \lsim 2 \times 10^{-2}
\eeq
In the case nonzero r.m.f. we obtain this is replaced by
\beq
\label{}
\sin^2 2\theta_m \lsim \frac{4 \times 10^{-2}}{x}
\eeq

Another complementary constraint can be obtained from the
requirement that in the non-trapping regime the sterile
neutrino can be emitted from anywhere inside the star
volume with a rate
$$
\frac{dQ(B=0)}{dt} \simeq \frac{4}{3}\pi R^3_{core}n_{\nu e}\Gamma_s\VEV{E_s}
\simeq 1.4\times 10^{55}\sin^2 2\theta_m \rm{\frac{J}{s}}
$$
which should not exceed the maximum observed integrated neutrino
luminosity. For instance, for the case of SN1987A, this is
$\sim 10^{46}~$ J, so that one obtains the excluded region
\cite{Maalampi}
\beq
\sin^2 2\theta_m \gsim 7 \times 10^{-10}
\label{limit4}
\eeq
In the case of a strong magnetic field $B \neq 0$
we use the known estimate for the active neutrino
collision rate \eq{B14} and the relationship between
the corresponding conversion probabilities in order
to obtain the ratio of sterile neutrino volume energy
losses in the presence and absence of magnetic field
\beq
\frac{dQ(B = 0)/dt}{dQ(B \neq 0)/dt} \sim \frac{1}{x B_{14}} ,
\label{limit6}
\eeq
where $x$ is the small parameter in \eq{argument}. From the
last inequality we can find a region of abundances where our
result for the conversion probability \eq{f11} is
valid ($x\ll 1$) so that we obtain the excluded region
\beq
\sin^2 2\theta_m \gsim \frac{7 \times 10^{-10}}{x B_{14}} \:.
\label{small}
\eeq
Note that this constraint on the \neu parameters
can be more stringent than that of \eq{limit4}.
In particular, for a supernova with strong magnetic field
it is possible to exclude all region of large mixing angles,
if the parameter $x$ in \eq{argument} is $x \leq 0.04$,
as we showed in Fig. 1 of ref. \cite{pastor}.
This will be realized for a r.m.s. field $B_{14} \sim 10^2$
\cite{Thomson} and 100 MeV mean sterile neutrino energy if the
abundance parameter is less than
\beq
\mid 3Y_e + 4Y_{\nu e} -1\mid \leq 0.3 \times
Y_e^{1/3}\rho_{14}^{-1/6}~.
\label{abun}
\eeq
This condition may indeed be realized for a stage of
supernova after bounce \cite{Maalampi,Notzold}. Moreover,
this assumption is not crucial for us, in contrast to the
case of resonant neutrino spin-flip due to a neutrino
magnetic moment.

\section{Conclusions}
\hspace{0.5cm}

The possible existence of huge random magnetic fields
that might be generated during the first few seconds of
neutrino emission in a supernova modifies the \neu
spectrum due to the magnetization of the medium, and thereby
affect the active to sterile neutrino conversion rates.
Their effect on the cooling rates may enable one to place
more stringent limits than those that apply in the absence
of a magnetic field.
This happens despite the fact that in the presence of
a large magnetic field the active to sterile \neu conversion
probability is suppressed relative to that in the zero field
case due to the larger energy difference between the two
diagonal entries in the \neu evolution hamiltonian
caused by the extra axial term. However, the sterile \neu
production rate could be larger in this case due to the effect
of the large magnetic field.
On the other hand the ratio of active and sterile \neu
thermal luminosities does not depend on the active \neu
production rate. However, the smaller the conversion
probability the larger the sterile \neu effective mean
free path, and therefore they can leave the star more easily
than in the case of zero magnetic field. This may lead to the
exclusion of the complete large mixing angle region \cite{pastor}.

\section*{Acknowledgements}
\hspace{0.5cm} This paper has been supported by DGICYT under
Grant number PB92-0084.


\Bibliography{15}

\bibitem{pa7}
 J. W.~F. Valle, \tp

\bibitem{BBNUTAU}
For a review see G. Steigman; proceedings of the
{\sl International School on Cosmological Dark Matter},
(World Scientific, 1994), ed. J. W. F. Valle and A. Perez, p. 55

\bibitem{SemikozValle}
V. Semikoz and J.W.F.Valle, \np{B425}{94}{651-664}.

\bibitem{Maalampi}
K. Kainulainen, J. Maalampi and J.T. Peltoniemi, \np{B358}{91}{435};
G. Raffelt and G. Sigl, \ap{1}{93}{165}.

\bibitem{pastor}
S. Pastor, V. Semikoz and J.W.F.Valle, FTUV/94-15,
hep-ph 9404299, \ap{}{94}{}, in press

\bibitem{Thomson}
C. Thomson and R.C. Dunkan, \apj{408}{93}{194}.

\bibitem{BFD}
J. Schechter and  J. W. F. Valle, \pr{D24}{81}{1883};
\pr{D25}{82}{283}

\bibitem{MSW}
M. Mikheyev, A. Smirnov, \sjnp{42}{86}{913};
L. Wolfenstein, \pr {D17}{78}{2369};\ib{D20}{79}{2634}.

\bibitem{SchrammB}
B. Cheng, D.N. Schramm and J.W. Truran, \pl{B316}{93}{521}.

\bibitem{Notzold}
D. Notzold \pr{D38}{88}{1658}

\end{thebibliography}
\end{document}